\documentclass[aps,prd,reprint,superscriptaddress,amsmath,amssymb,nofootinbib,floatfix]{revtex4-2}
\usepackage[utf8]{inputenc}
\usepackage[T1]{fontenc}
\usepackage{lmodern,bm,graphicx,booktabs,microtype,xcolor,tikz}
\usetikzlibrary{decorations.markings}
\usepackage[
  colorlinks=true,
  linkcolor=blue,
  citecolor=green!40!black,
  urlcolor=magenta!70!black
]{hyperref}
\graphicspath{{figures/}}
\newcommand{\MB}{M_B}
\usepackage{slashed}
\newcommand{\slsh}[1]{\slashed{#1}}
\newcommand{\dd}{\mathrm d}

\newcommand{\GeV}{\,\mathrm{GeV}}
\newcommand{\MeV}{\,\mathrm{MeV}}
\newcommand{\amp}{\mathcal M}

\begin{document}
\title{Nucleon-resonance contributions to the $p\bar{p}\to\eta_c\eta$ reaction}

\author{Wen-Ying Liu}
\email{liuwenying@lzu.edu.cn}
\affiliation{School of Physical Science and Technology, Lanzhou University, Lanzhou 730000, China}
\affiliation{Lanzhou Center for Theoretical Physics, Key Laboratory of Theoretical Physics of Gansu Province, Key Laboratory of Quantum Theory and Applications of MoE, Gansu Provincial Research Center for Basic Disciplines of Quantum Physics, Lanzhou University, Lanzhou 730000, China}
\affiliation{MoE Frontiers Science Center for Rare Isotopes, Lanzhou University, Lanzhou 730000, China}
\affiliation{Research Center for Hadron and CSR Physics, Lanzhou University and Institute of Modern Physics of CAS, Lanzhou 730000, China}
\author{Xiang Liu}
\email[Corresponding author: ]{xiangliu@lzu.edu.cn}
\affiliation{School of Physical Science and Technology, Lanzhou University, Lanzhou 730000, China}
\affiliation{Lanzhou Center for Theoretical Physics, Key Laboratory of Theoretical Physics of Gansu Province, Key Laboratory of Quantum Theory and Applications of MoE, Gansu Provincial Research Center for Basic Disciplines of Quantum Physics, Lanzhou University, Lanzhou 730000, China}
\affiliation{MoE Frontiers Science Center for Rare Isotopes, Lanzhou University, Lanzhou 730000, China}
\affiliation{Research Center for Hadron and CSR Physics, Lanzhou University and Institute of Modern Physics of CAS, Lanzhou 730000, China}

\begin{abstract}
Within an effective Lagrangian approach, we investigate the associated charmonium production reaction $p\bar{p}\to\eta_c\eta$ from threshold up to $\sqrt{s}=5.00\GeV$, incorporating intermediate contributions from the ground-state nucleon and the nucleon resonances $N(1535)$, $N(1650)$, $N(1895)$, and $N(1710)$. The effective resonance parameters are constrained by the BESIII measurement of the crossed decay $\eta_c\to p\bar{p}\eta$ for both the constructive (Solution~C) and destructive (Solution~D) radiative-interference solutions. Using these decay-constrained parameters, we predict the total and differential Born cross sections for the $p\bar{p}\to\eta_c\eta$ reaction. We show that coherent resonance exchanges dominate the near-threshold production dynamics and significantly enhance the total cross section compared to the pure nucleon-exchange contribution, which suffers from a strong dynamical cancellation. As the collision energy increases, the rapidly growing nucleon contribution progressively dilutes the resonance terms, reducing the relative difference between the total cross sections for Solutions~C and D. Furthermore, while the pure nucleon-exchange amplitude exhibits an exact dynamical zero at $\theta=90^\circ$ due to the pseudoscalar coupling structure, resonance exchanges fill this node in the predicted angular distributions, providing an observable test of the underlying exchange mechanisms. These predictions can be tested in future antiproton annihilation experiments at $\bar{\text{P}}\text{ANDA}$.
\end{abstract}
\maketitle

\section{Introduction}\label{sec:introduction}

Charmonium production in proton--antiproton ($p\bar{p}$) annihilation provides a useful setting for exploring nonperturbative quantum chromodynamics (QCD) and hadron spectroscopy~\cite{E760:1991xfp,E760:1992pgt,FermilabE835:2003ula,Andreotti:2005ts,PANDA:2009yku}. Unlike electron--positron ($e^+e^-$) annihilation, where direct single-photon formation is strictly limited to vector states ($J^{PC}=1^{--}$), $p\bar{p}$ collisions can directly access charmonia of all conventional spin and parity quantum numbers via multi-gluon intermediate states. Beyond direct $s$-channel formation, exclusive associated production reactions $p\bar{p}\to H_c X$ (with $H_c$ a charmonium state and $X$ a light meson) probe charmonium--baryon interactions, nonperturbative gluon dynamics, and hadron-exchange mechanisms.

Among such associated channels, the reaction $p\bar{p}\to\eta_c\eta$ holds a distinctive physical position. The outgoing $\eta$ meson is an isoscalar state ($I=0$), so strong isospin conservation at the $N^*N\eta$ vertex forbids the exchange of isospin-$3/2$ $\Delta$ resonances in the $t$ and $u$ channels. Consequently, the intermediate baryon spectrum is restricted entirely to nucleons and isospin-$1/2$ nucleon resonances ($N^*$). This feature isolates the coupling of pseudoscalar charmonium to baryon states without contamination from $\Delta$ excitations, providing a clean and direct theoretical counterpart to associated vector-charmonium production channels.

Experimental studies of exclusive associated charmonium production have advanced substantially over recent decades. Pioneering measurements by the E760 and E835 Collaborations at Fermilab demonstrated the feasibility of exploring exclusive charmonium channels with antiproton beams, including dedicated studies of the $p\bar{p}\to J/\psi\pi^0$ reaction during searches for the singlet state $h_c$~\cite{Armstrong:1992ae,Andreotti:2005vu}. More recently, the planned physics program of the $\bar{\text{P}}\text{ANDA}$ experiment at FAIR~\cite{PANDA:2009yku,PANDA:2021ozp} has renewed interest in systematic high-precision measurements of associated charmonium production using antiproton beams. On the theoretical front, various approaches have been developed to explore associated charmonium production across different kinematic domains. In early exploratory studies, Gaillard, Maiani, and Petronzio~\cite{Gaillard:1982zm} applied the soft-pion theorem to estimate the $p\bar{p}\to J/\psi\pi^0$ cross section near threshold. At sufficiently high energies, backward and forward production has been formulated by Pire et al.~\cite{Pire:2013jva} within collinear factorization in terms of baryon-to-meson transition distribution amplitudes (TDAs). For the near-threshold and intermediate-energy regimes accessible to antiproton beam facilities, the effective Lagrangian approach based on tree-level hadronic exchanges, formulated by Barnes and Li~\cite{Barnes:2006ck} for channels such as $p\bar{p}\to \pi^0\Psi$ and subsequently applied to $\eta$ associated production~\cite{Lin:2012ru,Zhou:2020smj}, provides a practical and transparent phenomenological framework based on $t$- and $u$-channel baryon exchanges.

Within the effective Lagrangian framework, the exclusive associated charmonium production process $p\bar{p}\to H_c X$ is directly linked to the crossed three-body decay $H_c\to p\bar{p} X$ via crossing symmetry~\cite{Lundborg:2005am,Barnes:2006ck,Barnes:2010yb}. Charmonium three-body decays provide valuable empirical information on baryon excitations and effective charmonium--baryon couplings~\cite{Zou:2008be,Liu:2005pm}. Early phenomenological calculations frequently relied on pure
ground-state nucleon exchange supplemented by empirical form factors~\cite{Liang:2004sd,Lin:2012ru,Barnes:2007ub}. However, subsequent investigations revealed that intermediate nucleon resonances ($N^*$) and their coherent interference can substantially alter both the absolute magnitude of cross sections and the shapes of angular distributions~\cite{VandeWiele:2013dli,Xu:2015qqa,Wang:2017sxq,Zhou:2020smj}. For vector charmonia, analyses of $e^+e^-\to p\bar{p}\pi^0$ and related decay spectra demonstrated that intermediate $N^*$ excitations play an important role in reactions such as $p\bar{p}\to\psi(3770)\pi^0$ and $p\bar{p}\to Y(4220)\pi^0$~\cite{Xu:2015qqa,Wang:2017sxq}. More recently, an effective Lagrangian study of $p\bar{p}\to\psi\eta$ [$\psi=J/\psi,\psi(3686)$] constrained by $\psi\to p\bar{p}\eta$ decay data showed that intermediate $N^*$ resonances can substantially modify the production observables~\cite{Zhou:2020smj}.

In contrast to these developments in the vector-charmonium sector, the pseudoscalar reaction $p\bar{p}\to\eta_c\eta$ has remained poorly understood. Without data on the crossed decay $\eta_c\to p\bar{p}\eta$, earlier calculations relied solely on ground-state nucleon exchange~\cite{Lin:2012ru}. This experimental limitation has been overcome by the BESIII Collaboration, which reported the first observation of the three-body decay $\eta_c\to p\bar{p}\eta$ via the radiative transition $\psi(3686)\to\gamma\eta_c$ with a statistical significance exceeding $10\sigma$~\cite{BESIII:2026kdm}. The measured invariant-mass distributions of the $p\eta$ and $\bar{p}\eta$ subsystems display prominent low-mass enhancements and broad intermediate structures, motivating a comprehensive description incorporating intermediate $N^*$ contributions. Moreover, due to quantum interference between the resonant $M1$ transition and non-resonant background, the BESIII measurement yields two acceptable branching-fraction solutions: constructive interference (Solution~C) and destructive interference (Solution~D). Under crossing symmetry, the decay $\eta_c\to p\bar{p}\eta$ and the production reaction $p\bar{p}\to\eta_c\eta$ share the same effective hadronic vertices. The published BESIII invariant-mass projections ($M_{p\eta}$, $M_{\bar{p}\eta}$, and $M_{p\bar{p}}$) thus provide essential empirical constraints on the effective coupling products and relative strong phases.

Motivated by these experimental results, we carry out a phenomenological study of the associated production reaction $p\bar{p}\to\eta_c\eta$ within an effective Lagrangian framework constrained by the BESIII decay measurements. We formulate a dynamical model basis, $\MB$, which incorporates coherent $t$- and $u$-channel exchanges of the ground-state nucleon $N(939)$ alongside four established spin-$1/2$ resonances: the negative-parity states $N(1535)$, $N(1650)$, and $N(1895)$, and the positive-parity state $N(1710)$. The ground-state nucleon coupling is fixed using the empirical $\eta NN$ coupling and the experimental $\eta_c\to p\bar{p}$ partial width. The effective coupling-product moduli and relative phases of the four nucleon resonances are extracted via simultaneous fits to the BESIII invariant-mass projections for both Solution~C and Solution~D. Continuing these constrained amplitudes to the production region, we calculate the Born total cross sections, analyze the competition between nucleon and resonance exchanges, and examine the resulting energy-dependent difference between Solutions~C and D. We systematically investigate the sensitivity of the predicted cross sections to variations of the form-factor cutoff parameter $\beta$ ($2\leq\beta\leq6$), and assess the potential for future $p\bar{p}$ annihilation experiments to test the two radiative-interference solutions in light of parameter uncertainties. Furthermore, we compute both absolute and normalized angular distributions, demonstrating that coherent resonance exchanges fill the dynamical dip of the nucleon pole at central angles.

The paper is organized as follows. In Sec.~\ref{sec:framework}, we establish the effective Lagrangians, reaction kinematics, and tree-level baryon-exchange amplitudes for $p\bar{p}\to\eta_c\eta$. In Sec.~\ref{sec:decay-constraints}, we formulate the crossed $\eta_c\to p\bar{p}\eta$ decay amplitudes, outline the physical criteria for selecting the resonance basis, and present simultaneous fits to the BESIII data. In Sec.~\ref{sec:production-results}, we report numerical predictions for total Born cross sections, examine the underlying reaction mechanisms and angular distributions, and assess model sensitivities. A summary and outlook are presented in Sec.~\ref{sec:summary}. Technical details regarding data conversion, the fitting objective function, and uncertainty propagation are collected in Appendix~\ref{app:experimental-input}.

\section{Production mechanism of \texorpdfstring{$p\bar p\to\eta_c\eta$}{ppbar to eta-c eta}}\label{sec:framework}

\begin{figure}[!htbp]
\centering
\includegraphics[width=\columnwidth]{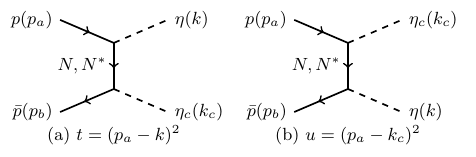}
\caption{Baryon-exchange diagrams describing the $p\bar p\to\eta_c\eta$ reaction: (a) $t$-channel exchange and (b) $u$-channel exchange. Solid and dashed lines denote baryons and mesons, respectively.}\label{fig:baryons}
\end{figure}

We investigate the associated charmonium production process
\begin{equation}
 p(p_a)\,\bar p(p_b)\to\eta_c(k_c)\,\eta(k).\label{eq:production}
\end{equation}
The external four-momenta define the standard Mandelstam variables and intermediate transfer momenta:
\begin{equation}
\begin{gathered}
 q_t=p_a-k, \quad
q_u=p_a-k_c, \\
t=q_t^2, \quad
u=q_u^2, \quad
s=(p_a+p_b)^2.
\end{gathered}
\end{equation}
At tree level, the reaction proceeds via the exchange of intermediate nucleons and nucleon excited states in the $t$ and $u$ channels, as illustrated in Fig.~\ref{fig:baryons}.

In this work, we consider the following baryon-exchange basis $\MB$ for the production process:
\begin{equation}
 \begin{aligned}\MB:\;&N(939)\,(\frac{1}{2}^+) + N(1710)\,(\frac{1}{2}^+) \\&+ N(1650)\,(\frac{1}{2}^-) + N(1895)\,(\frac{1}{2}^-) \\
 &+ N(1535)\,(\frac{1}{2}^-).
 \end{aligned}\label{eq:baseline}
\end{equation}
The exchanged baryons are far off shell in the production reaction, motivating our emphasis on low-lying nucleon excitations with smaller virtualities. Because the reaction involves the $N^*N\eta$ vertex, we also favor resonances with sizable $N\eta$ branching fractions. A detailed discussion of the resonance selection criteria is presented in Sec.~\ref{subsec:decay-amplitudes}.

The effective interaction Lagrangians for the pseudoscalar mesons $\phi\in\{\eta,\eta_c\}$ are given by
\begin{align}
 \mathcal L_{\phi NN}&=-i g_{\phi NN}\bar N\gamma_5\phi N,\label{eq:lagN}\\
 \mathcal L_{\phi NR_-}&=-g_{\phi NR_-}\bar N\phi R_-+\mathrm{H.c.},\label{eq:lagS}\\
 \mathcal L_{\phi NR_+}&=-\frac{g_{\phi NR_+}}{2m_N}\bar N\gamma_5\gamma^\mu(\partial_\mu\phi)R_++\mathrm{H.c.},\label{eq:lagP}
\end{align}
where $R_-$ denotes the three negative-parity states [$N(1535)$, $N(1650)$, and $N(1895)$] and $R_+$ represents the positive-parity state $N(1710)$. The pseudoscalar $\eta NN$ and nonderivative scalar-type $\eta NR_-$ couplings follow Ref.~\cite{Zhou:2020smj}, while the $\eta_cNN$ interaction is taken from Ref.~\cite{Lin:2012ru}. For the positive-parity spin-$1/2$ resonance, we adopt the derivative pseudoscalar-meson coupling structure used for the $\pi N P_{11}$ vertex in Ref.~\cite{Wang:2017sxq}, extended here to the $\eta N R_+$ vertex. By parity conservation, these interaction structures extend directly to the pseudoscalar charmonium vertices $\eta_c NR_\pm$.

In the center-of-mass frame of $p\bar p\to\eta_c\eta$, let $\theta\equiv\theta_\eta$ denote the scattering angle between the outgoing $\eta$ meson and the incoming proton. The magnitudes of the initial and final three-momenta are $p_i=\sqrt{s/4-m_N^2}$ and $p_f=\lambda^{1/2}(s,m_{\eta_c}^2,m_\eta^2)/(2\sqrt{s})$, where $\lambda(a,b,c)=(a-b-c)^2-4bc$ is the K\"all\'en function. The kinematic limits of the Mandelstam variables $t$ and $u$ are identical, $u_\pm(s)=t_\pm(s)$, with
\begin{equation}
 t_\pm(s)=m_N^2+\frac{m_\eta^2+m_{\eta_c}^2-s}{2}\pm2p_ip_f.
 \label{eq:virtuality}
\end{equation}
Throughout the physical phase space of $p\bar{p}\to\eta_c\eta$, the momentum transfers remain strictly subthreshold, satisfying $t,u\leq(m_N-m_\eta)^2$. Both $t$ and $u$ thus lie below the $N\pi$ threshold, the lowest strong-decay threshold for the exchanged baryons. The corresponding self energies have no absorptive part in this region. At tree level we use the real propagator denominators $t-m_i^2$ and $u-m_i^2$.

Applying the Feynman rules derived from Eqs.~\eqref{eq:lagN}--\eqref{eq:lagP} to Fig.~\ref{fig:baryons}, the component scattering amplitudes for nucleon and resonance exchanges are given by
\begin{align}
 \amp_N^p={}&C_N\bar v(p_b)\Bigg[
 \gamma_5(\slsh q_t+m_N)\gamma_5\frac{F_N(t)}{t-m_N^2}\nonumber\\
 &+\gamma_5(\slsh q_u+m_N)\gamma_5\frac{F_N(u)}{u-m_N^2}\Bigg]u(p_a),\label{eq:crossN} \displaybreak[1]\\
 \amp_{R_-}^p={}&C_{R_-}\bar v(p_b)\Bigg[
 (\slsh q_t+m_R)\frac{F_R(t)}{t-m_R^2}\nonumber\\
 &+(\slsh q_u+m_R)\frac{F_R(u)}{u-m_R^2}\Bigg]u(p_a),\label{eq:crossS} \displaybreak[1]\\
 \amp_{R_+}^p={}&-\frac{C_{R_+}}{(2m_N)^2}\bar v(p_b)\Bigg[\nonumber\\
 &\slsh k_c\gamma_5(\slsh q_t+m_R)\gamma_5\slsh k\frac{F_R(t)}{t-m_R^2}\nonumber\\
 &\hspace{2mm}+\gamma_5\slsh k(\slsh q_u+m_R)\slsh k_c\gamma_5\frac{F_R(u)}{u-m_R^2}\Bigg]u(p_a),\label{eq:crossP}
\end{align}
where the first and second terms in each bracket correspond to $t$- and $u$-channel exchanges, respectively. To account for the extended internal structure of hadrons and regularize the high-energy behavior, each exchanged baryon is dressed with an off-shell form factor~\cite{Wang:2017sxq,Zhou:2020smj}:
\begin{equation}
 F_i(z)=\frac{\Lambda_i^4}{\Lambda_i^4+(z-m_i^2)^2},\label{eq:FF}
\end{equation}
where $\Lambda_i=m_i+\beta\Lambda_{\rm QCD}$ with  $\Lambda_{\rm QCD}=0.22\GeV$ and $z$ denotes the squared momentum transfer ($t$ or $u$), normalized such that $F_i(m_i^2)=1$ on shell. The dimensionless parameter $\beta$ controls the cutoff scale, with larger values corresponding to softer off-shell suppression. A common value of $\beta$ is adopted for all intermediate baryon lines. Although $\beta$ could in principle be treated as an adjustable parameter in the fit, the present $\eta_c\to p\bar{p}\eta$ decay spectra provide only weak constraints on it. In the decay channel, the dominant contributions come from intermediate baryons near their mass shells, where $F_i\approx 1$; the decay spectra are therefore weakly sensitive to $\beta$, which mainly controls off-shell damping. We therefore adopt $\beta=4$ as our reference value, consistent with the phenomenological result $\beta=3.70\pm0.82$ obtained in $\psi(3686)\to p\bar{p}\eta$~\cite{Zhou:2020smj}. To assess the cutoff sensitivity, we vary $\beta$ over $2\leq\beta\leq6$ in Sec.~\ref{subsec:total-cross-sections}, refitting the decay parameters at each value of $\beta$.

The dimensionless tree-level coupling products appearing in Eqs.~\eqref{eq:crossN}--\eqref{eq:crossP} are defined as
\begin{align}
 C_N&=g_{\eta NN}g_{\eta_cNN},\qquad C_{R_-}=-g_{\eta NR_-}g_{\eta_cNR_-},\nonumber\\
 C_{R_+}&=g_{\eta NR_+}g_{\eta_cNR_+}.\label{eq:coefficients}
\end{align}
The total coherent production amplitude is constructed as~\cite{Zhou:2020smj}
\begin{equation}
 \amp_p=\amp_N^p+\sum_R\amp_R^pe^{+i\phi_R},
 \label{eq:Mp}
\end{equation}
where $R\in\{N(1535), N(1650), N(1895), N(1710)\}$. The relative phases $\phi_R$ parameterize effective strong-phase differences generated by rescattering and loop corrections beyond tree level. Without loss of generality, we set $C_i\geq0$ by absorbing any overall sign into the corresponding phase $\phi_i$, since the joint transformation $C_i\to-C_i$ and $\phi_i\to\phi_i+\pi$ leaves the amplitude invariant. The nucleon pole serves as the reference amplitude, with $C_N>0$ and $\phi_N=0$.

We fix the nucleon coupling product $C_N$ using the established $\eta NN$ coupling $g_{\eta NN}=8.99$~\cite{Zhou:2020smj} and the measured $\eta_c\to p\bar p$ partial width. The charmonium--nucleon coupling $g_{\eta_cNN}$ is obtained from the tree-level decay width relation:
\begin{equation}
 \Gamma(\eta_c\to p\bar p)=\frac{g_{\eta_cNN}^2 m_{\eta_c}}{8\pi}
 \sqrt{1-\frac{4m_N^2}{m_{\eta_c}^2}}.\label{eq:CN}
\end{equation}
Using the branching fraction $\mathcal B(\eta_c\to p\bar p)=(1.11\pm0.12)\times10^{-3}$ from the 2026 pdgLive listing~\cite{PDG:2026etacOnline} as a fixed input, we obtain $g_{\eta_cNN}\simeq0.01915$, which yields the positive real reference value
\begin{equation}
 C_N=g_{\eta NN}g_{\eta_cNN}\simeq17.22\times10^{-2}.
\end{equation}

The differential and integrated total Born cross sections are given by
\begin{equation}
\begin{gathered}
 \frac{\dd\sigma}{\dd\cos\theta}=\frac{p_f}{32\pi s\,p_i}\overline{|\amp_p|^2}, \quad
 \sigma(s)=\int_{-1}^{1}\dd\cos\theta\,\frac{\dd\sigma}{\dd\cos\theta},\label{eq:crosssection}
\end{gathered}
\end{equation}
where $\overline{|\amp_p|^2}$ denotes the spin-averaged squared amplitude.

The fixed hadron masses and widths adopted throughout this work are summarized in Table~\ref{tab:inputs}. The resonance vertices enter the amplitudes through the products $C_R$. The available data constrain these products but do not determine the individual meson--baryon couplings separately. To constrain these products and their relative phases, we analyze the crossed decay $\eta_c\to p\bar p\eta$.

\begin{table}[htbp]\centering
\caption{Hadron masses and widths adopted in this work, taken from the PDG~\cite{ParticleDataGroup:2024cfk}. Masses and widths are listed in MeV.}\label{tab:inputs}
\renewcommand{\arraystretch}{1.4}
\begin{tabular*}{0.7\columnwidth}{@{\extracolsep{\fill}}lcc@{}}\toprule
State or quantity & Mass & Width \\\midrule
$\eta_c$ & 2984.1 & 30.5 \\
$N(939)$ & 938.3 & -- \\
$\eta$ & 547.9 & -- \\
$N(1535)$ & 1530.0 & 150.0 \\
$N(1650)$ & 1650.0 & 125.0 \\
$N(1895)$ & 1895.0 & 120.0 \\
$N(1710)$ & 1710.0 & 140.0 \\
\bottomrule
\end{tabular*}
\end{table}

\section{Determination of model parameters from \texorpdfstring{$\eta_c\to p\bar p\eta$}{eta-c decay}}\label{sec:decay-constraints}

\subsection{Decay amplitude and relation to production}\label{subsec:decay-amplitudes}

\begin{figure}[!htbp]
\centering
\includegraphics[width=\columnwidth]{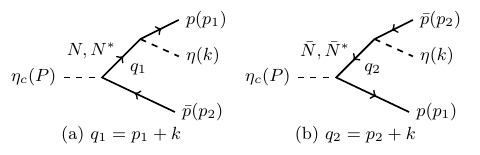}
\caption{Baryon-exchange diagrams describing the decay
$\eta_c\to p\bar p\eta$:
(a) intermediate baryon decaying into $p\eta$, and
(b) intermediate antibaryon decaying into $\bar p\eta$.}
\label{fig:decay-baryons}
\end{figure}

To constrain the effective coupling products and relative phases entering the production amplitude, we analyze the three-body decay process
\begin{equation}
 \eta_c(P)\to p(p_1)\,\bar p(p_2)\,\eta(k).\label{eq:decay}
\end{equation}
The tree-level Feynman diagrams are shown in Fig.~\ref{fig:decay-baryons}. Crossing symmetry relates the two reactions through the same effective vertices~\cite{Lundborg:2005am,Barnes:2010yb}. However, while intermediate virtualities in the production channel are subthreshold, the timelike propagators in the decay channel directly probe the physical resonance region within the kinematic boundaries
\begin{equation}
\begin{gathered}
 1.49\leq M_{p\eta},M_{\bar p\eta}\leq2.05\GeV,
 \\
 1.88\leq M_{p\bar p}\leq2.44\GeV.\label{eq:windows}
\end{gathered}
\end{equation}

In the experimental measurement of the decay $\eta_c\to p\bar{p}\eta$ by the BESIII Collaboration~\cite{BESIII:2026kdm}, the invariant-mass distributions of the $p\eta$ and $\bar{p}\eta$ subsystems display prominent deviations from the pure phase-space expectation, marked by conspicuous low-mass enhancements near $1.5\GeV$. In their analysis, intermediate nucleon states were introduced to parameterize these structures for efficiency modeling, motivating a theoretical description incorporating intermediate nucleon resonances in the $1.5\GeV$ mass region. To establish a physically grounded resonance basis $\MB$, we examine established nucleon resonances listed by the Particle Data Group (PDG)~\cite{ParticleDataGroup:2024cfk} within the kinematically accessible decay window ($1.49\leq M_{p\eta}\leq2.05\GeV$). We prioritize four-star resonances with sizable $N\eta$ branching fractions and low-orbital-angular-momentum couplings to limit centrifugal suppression.

Guided by these physical criteria, we adopt a resonance basis $\MB$ consisting of four well-established four-star spin-$1/2$ resonances: $N(1535)$, $N(1650)$, $N(1895)$, and $N(1710)$. The three negative-parity states $N(1535)$, $N(1650)$, and $N(1895)$ ($J^P=1/2^-$) couple to $N\eta$ in an $S$ wave with substantial branching fractions ($\mathcal{B}\sim 15\%$--$55\%$), while the positive-parity state $N(1710)$ ($J^P=1/2^+$) couples in a $P$ wave with $\mathcal{B}\sim 10\%$--$50\%$~\cite{ParticleDataGroup:2024cfk,Klempt:2009pi,Crede:2013kia,Manley:1992yb,Shrestha:2012ep,Hunt:2018wqz}. Other known states are not included in this model: although the three-star state $N(1880)\,(1/2^+)$ shares the $P$-wave coupling feature, its properties remain less constrained empirically; the Roper resonance $N(1440)\,(1/2^+)$ has its nominal mass below the physical $N\eta$ threshold ($1.486\GeV$), and the potential off-shell contribution of this subthreshold state is not explicitly parameterized here; states such as $N(1520)\,(3/2^-)$, $N(1675)\,(5/2^-)$, and $N(1680)\,(5/2^+)$ exhibit negligible $N\eta$ branches ($\lesssim 1\%$) and higher partial waves; and while the positive-parity spin-$3/2$ states $N(1720)\,(3/2^+)$ and $N(1900)\,(3/2^+)$ have modest branching fractions ($1\%$--$14\%$), their production in $\eta_c\to\bar{N}N^*$ requires a ${}^5D_0$ ($L=2$) orbital configuration, which is suppressed near threshold by the centrifugal barrier. The basis $\MB$ describes the low-partial-wave contributions retained here. The uncertainty from omitted resonances has not been quantified.

In the decay kinematics, we define the intermediate four-momenta as $q_1=p_1+k$ and $q_2=p_2+k$, with invariant masses squared $s_j=q_j^2$ ($j=1,2$). The hadronic form factors $F_i(s_j)$ retain the functional form of Eq.~\eqref{eq:FF}. The denominator for the stable ground-state nucleon is $D_N(z)=z-m_N^2$, whereas intermediate resonances are parameterized using relativistic Breit--Wigner propagators:
\begin{equation}
 D_R(z)=z-m_R^2+i m_R\Gamma_R(z).\label{eq:den}
\end{equation}
For $N(1650)$, $N(1895)$, and $N(1710)$, we use constant total widths $\Gamma_R(z)\equiv\Gamma_R$ from Table~\ref{tab:inputs}. For $N(1535)$, which lies in close proximity to the $N\eta$ threshold, we adopt an energy-dependent width parametrization motivated by Ref.~\cite{Zhou:2020smj}:
\begin{equation}
 \Gamma_{1535}(z)=\Gamma_0\big[0.42\,r_\pi(z)+0.42\,r_\eta(z)+0.16\big],\label{eq:running}
\end{equation}
where $0.42$, $0.42$, and $0.16$ are the adopted nominal branching fractions to $N\pi$, $N\eta$, and remaining inelastic channels, respectively, with $\Gamma_0=150\MeV$. The kinematic factors $r_a(z)$ ($a=\pi,\eta$) are given by
\begin{align}
 r_a(z)&=\frac{q_a(z)}{q_a(m_R^2)}
 \frac{[E_N(z)+m_N]/\sqrt z}{[E_N(m_R^2)+m_N]/m_R},\\
 q_a(z)&=\frac{\lambda^{1/2}(z,m_N^2,m_a^2)}{2\sqrt z},\qquad
 E_N(z)=\sqrt{m_N^2+q_a(z)^2}.
\end{align}
Below two-body threshold, $z<(m_N+m_a)^2$, we set $q_a(z)=0$ in the corresponding partial width, keeping the inelastic term constant.

The component decay amplitudes are explicitly expressed as
\begin{widetext}
\begin{align}
 \amp_N^d={}&C_N\bar u(p_1)\Bigg[
 \gamma_5(\slsh q_1+m_N)\gamma_5\frac{F_N(s_1)}{s_1-m_N^2}
 +\gamma_5(-\slsh q_2+m_N)\gamma_5\frac{F_N(s_2)}{s_2-m_N^2}\Bigg]v(p_2),\label{eq:pair}\\
 \amp_{R_-}^d={}&C_{R_-}\bar u(p_1)\Bigg[
 (\slsh q_1+m_R)\frac{F_R(s_1)}{s_1-m_R^2+i m_R\Gamma_R(s_1)} + (-\slsh q_2+m_R)\frac{F_R(s_2)}{s_2-m_R^2+i m_R\Gamma_R(s_2)}\Bigg]v(p_2),\label{eq:Bminus} \\
 \amp_{R_+}^d={}&\frac{C_{R_+}}{(2m_N)^2}\bar u(p_1)\Bigg[
 \gamma_5\slsh k(\slsh q_1+m_R)\slsh P\gamma_5\frac{F_R(s_1)}{s_1-m_R^2+i m_R\Gamma_R(s_1)}\nonumber\\
 &\hspace{20mm}+\gamma_5\slsh P(-\slsh q_2+m_R)\slsh k\gamma_5\frac{F_R(s_2)}{s_2-m_R^2+i m_R\Gamma_R(s_2)}\Bigg]v(p_2).\label{eq:Bplus}
\end{align}
\end{widetext}
The total coherent decay amplitude is given by
\begin{equation}
 \amp_d=\amp_N^d+\sum_R\amp_R^de^{+i\phi_R},
 \label{eq:Md}
\end{equation}
with $R\in\{N(1535), N(1650), N(1895), N(1710)\}$.

\begin{figure*}[!htp]\centering
\includegraphics[width=0.7\textwidth]{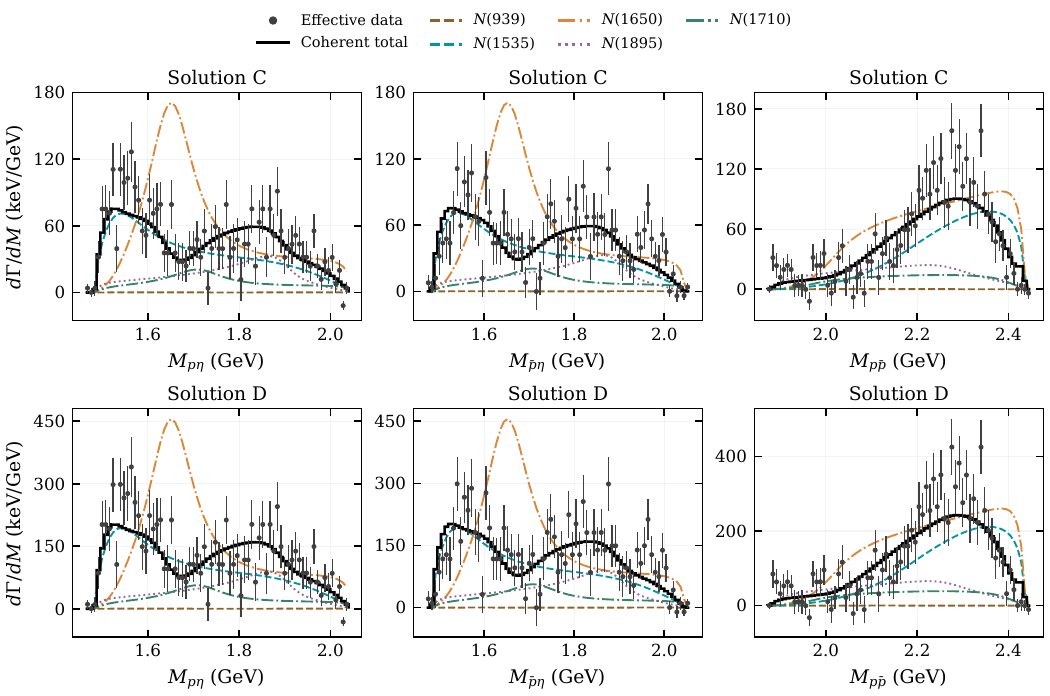}
\caption{Simultaneous fits of model $\MB$ to the three two-body decay invariant-mass projections for Solution~C (upper panels) and Solution~D (lower panels). Data points with error bars represent the effective background-subtracted spectra; solid black histograms denote the bin-integrated theoretical predictions. Dashed and dash-dotted curves illustrate the isolated contributions of individual intermediate exchanges. All curves are evaluated at the best-fit parameters for $\beta=4$.}\label{fig:fitMB}
\end{figure*}

The kinematic structures of the decay and production amplitudes are related under crossing symmetry by the momentum substitutions
\begin{equation}
 p_1\to-p_b,\qquad p_2\to-p_a,\qquad P\to-k_c,
\end{equation}
with the outgoing $\eta$ momentum $k$ unchanged, yielding $q_1\to q_u$ and $q_2\to-q_t$. While crossing symmetry directly connects the tree-level amplitude structures under the kinematic momentum replacements, the continuation of fitted effective phases across crossed channels is not guaranteed by symmetry alone. In the timelike decay domain, these effective phases can absorb channel-dependent final-state interactions and coupled-channel unitarity cuts~\cite{Dedonder:2018ulo}. Here we retain the fitted phases as energy-independent phenomenological parameters~\cite{Zhou:2020smj}.

For a two-body subsystem $(ab)$ with invariant mass $m$ and spectator particle $c$, the breakup momentum of $c$ in the parent rest frame and that of $a$ (or $b$) in the $(ab)$ rest frame are given by
\begin{equation}
\begin{gathered}
 k_o=\frac{\lambda^{1/2}(M^2,m^2,m_c^2)}{2M}, \quad
 k_i=\frac{\lambda^{1/2}(m^2,m_a^2,m_b^2)}{2m},
\end{gathered}
\end{equation}
The theoretical differential decay rate and bin-integrated partial width read
\begin{equation}
\begin{gathered}
 \frac{\dd\Gamma}{\dd m}=\frac{k_ok_i}{64\pi^3M^2}
 \int_{-1}^{1}\dd z\,\overline{|\amp_d|^2},\\
 \Delta\Gamma_{\alpha b}^{\rm th}=\int_{m_b^-}^{m_b^+}\dd m\,
 \frac{\dd\Gamma}{\dd m},\label{eq:spectrum}
\end{gathered}
\end{equation}
where $z$ is the cosine of the helicity angle, $[m_b^-, m_b^+]$ denotes the bin interval for projection $\alpha \in \{p\eta, \bar{p}\eta, p\bar{p}\}$, and $\overline{|\amp_d|^2}\equiv\sum_{\text{spins}}|\amp_d|^2$ denotes the sum over final-state baryon spins with no initial spin average since $\eta_c$ is spinless.

\begin{table}[htbp]\centering
\caption{Fitted coupling-product moduli and relative phases in model $\MB$, with conditional MINOS uncertainties at central cutoff $\beta=4$. Phases are in degrees.}
\label{tab:MB}
\small\setlength{\tabcolsep}{2pt}
\renewcommand{\arraystretch}{1.4}
\resizebox{\columnwidth}{!}{
\begin{tabular}{crrrr}\toprule
Component & \multicolumn{1}{c}{$10^2C_i$ (C)} & \multicolumn{1}{c}{$\phi_i$ (C)} & \multicolumn{1}{c}{$10^2C_i$ (D)} & \multicolumn{1}{c}{$\phi_i$ (D)}\\\midrule
$N(939)$  & \multicolumn{1}{c}{17.22} & \multicolumn{1}{c}{0.00} & \multicolumn{1}{c}{17.22} & \multicolumn{1}{c}{0.00}\\
$N(1535)$ & $7.64{}^{+0.73}_{-0.69}$ & $-100.30{}^{+35.73}_{-36.08}$ & $12.62{}^{+1.18}_{-1.11}$ & $-100.88{}^{+46.07}_{-46.50}$\\
$N(1650)$ & $7.23{}^{+0.53}_{-0.55}$ & $104.56{}^{+34.25}_{-34.93}$ & $11.81{}^{+0.87}_{-0.90}$ & $103.68{}^{+44.91}_{-45.66}$\\
$N(1895)$ & $5.28{}^{+0.56}_{-0.60}$ & $15.87{}^{+38.41}_{-39.43}$ & $8.72{}^{+0.91}_{-0.98}$ & $14.58{}^{+48.01}_{-49.10}$\\
$N(1710)$ & $4.11{}^{+0.55}_{-0.60}$ & $-156.70{}^{+35.41}_{-37.80}$ & $6.78{}^{+0.85}_{-0.94}$ & $-159.22{}^{+45.73}_{-48.11}$\\
\bottomrule\end{tabular}}
\end{table}

\subsection{Fit results and resonance contributions}\label{subsec:decay-fit-results}

Before presenting the fit results, we briefly outline the data conversion and fitting strategy, with full technical details provided in Appendix~\ref{app:experimental-input}. In the BESIII measurement of $\psi(3686)\to\gamma\eta_c\to\gamma p\bar{p}\eta$~\cite{BESIII:2026kdm}, quantum interference between the resonant $M1$ radiative transition and the non-resonant continuum yields two distinct branching-fraction normalizations: Solution~C (constructive interference) and Solution~D (destructive interference). For each solution, the experimental background-subtracted event yields across the three invariant-mass projections ($M_{p\eta}$, $M_{\bar{p}\eta}$, and $M_{p\bar{p}}$) are converted into absolute differential decay widths $(\dd\Gamma/\dd m)$. We then evaluate the corresponding theoretical differential distributions using the coherent decay amplitude in Eq.~\eqref{eq:Md} and the bin-integrated partial width formula in Eq.~\eqref{eq:spectrum}. By minimizing the composite $\chi^2$ function simultaneously over all three projections, we extract the effective coupling products $C_R$ and relative strong phases $\phi_R$ for Solutions~C and D independently.

The simultaneous fits to the BESIII projections are compared with experimental data in Fig.~\ref{fig:fitMB}, and the extracted coupling products and relative phases are summarized in Table~\ref{tab:MB}. The model describes the main features of the three invariant-mass projections. In particular, the low-mass enhancement near $1.53\GeV$ in the $p\eta$ and $\bar p\eta$ spectra is naturally accommodated through the coherent interplay of the $N(1535)$ and $N(1650)$ resonances. Although the isolated $N(1650)$ contribution is large, destructive interference with $N(1535)$ moderates the net near-threshold yield. The overall lineshape is also affected by the coherent contributions of the heavier states $N(1895)$ and $N(1710)$, which populate the broader intermediate-mass structures and their kinematic reflections in the $M_{p\bar p}$ spectrum.

Using the same independently fixed nucleon coupling $C_N$ in both solutions breaks exact scaling with the experimental branching fraction, so Solutions~C and D must be fitted separately. Since the off-shell nucleon contribution is negligible in the decay region, the fitted resonance moduli approximately scale as $\sqrt{\mathcal B_D/\mathcal B_C}\simeq1.64$, while the central relative phases are virtually identical between the two solutions. Although individual phases exhibit correlated uncertainties of several tens of degrees, the relative phase difference between the two $S_{11}$ states is tightly constrained: local covariance estimates give $\phi_{1650}-\phi_{1535}=(-155.14\pm6.56)^\circ$ for Solution~C and $(-155.44\pm6.51)^\circ$ for Solution~D. This indicates that the experimental decay distributions predominantly constrain the mutual interference between the two low-lying $S_{11}$ resonances. The fits yield $\chi^2_{\rm comp}/n_{\rm dof}=1.53$ and $1.54$ for Solutions~C and D, respectively. 

\section{Numerical results and discussion }\label{sec:production-results}

\subsection{Total cross sections}\label{subsec:total-cross-sections}

\begin{figure}[!tbp]\centering
\includegraphics[width=.8\columnwidth]{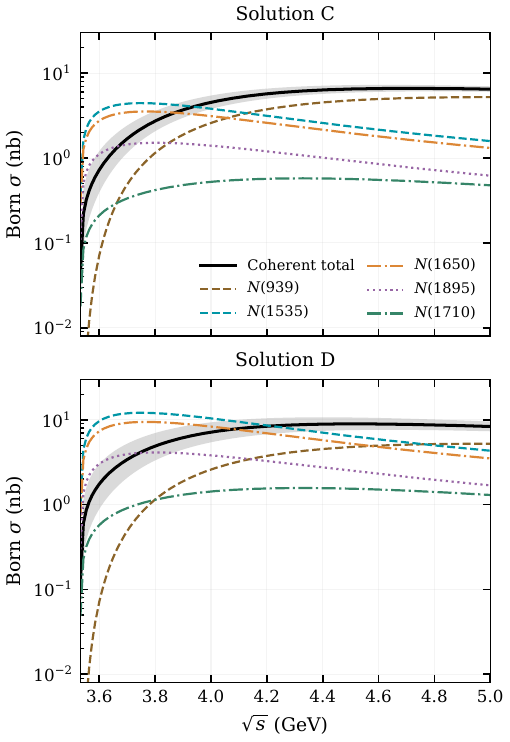}
\caption{Born total cross sections for $p\bar{p}\to\eta_c\eta$ as a function of the center-of-mass energy $\sqrt{s}$ at central cutoff $\beta=4$ for Solution~C (upper panel) and Solution~D (lower panel). The solid black curves show the coherent total cross section, with gray shaded bands denoting the pointwise nonlinear conditional fit uncertainties at $\Delta\chi^2_{\rm comp}=1$. Dashed and dash-dotted curves represent isolated contributions from individual baryon exchanges.}\label{fig:componentsMB}
\end{figure}

In this section, we substitute the effective resonance coupling products and relative phases determined from the decay $\eta_c\to p\bar{p}\eta$ (summarized in Table~\ref{tab:MB}) into the production amplitude in Eq.~\eqref{eq:Mp} using the crossed tree-level structures and the constant-phase continuation prescription specified in Sec.~\ref{sec:decay-constraints}. The differential and integrated total Born cross sections for $p\bar{p}\to\eta_c\eta$ are then evaluated using Eq.~\eqref{eq:crosssection}.

Figure~\ref{fig:componentsMB} displays the energy dependence of the predicted Born cross sections evaluated at the central form-factor cutoff $\beta=4$. The coherent total cross section and individual exchange contributions are calculated using the best-fit parameters, with shaded bands denoting the conditional uncertainty ranges propagated from the decay fits. The corresponding numerical values at representative center-of-mass energies, along with the conditional parameter uncertainties (the first error) and the cutoff sensitivity ranges across $2\leq\beta\leq6$ (the second error), are compiled in Table~\ref{tab:crossMB}.

\begin{table}[!htbp]\centering
\caption{Born total cross sections (in nb) at representative center-of-mass energies evaluated at the central cutoff $\beta=4$. The first uncertainty ($\mathrm{fit}$) corresponds to the conditional parameter profile at $\Delta\chi^2_{\rm comp}=1$, while the second ($\beta$) reflects the sensitivity envelope over the refitted cutoff scan $2\leq\beta\leq6$. 
}\label{tab:crossMB}
\small\setlength{\tabcolsep}{2pt}
\renewcommand{\arraystretch}{1.4}
\resizebox{\columnwidth}{!}{
\begin{tabular}{lrrr}
\toprule
$\sqrt{s}$ (GeV) & $N(939)$ & $\MB$, Solution~C & $\MB$, Solution~D\\
\midrule
3.60 & $0.0715{}^{+0.0001}_{-0.0110}\big|_\beta$ & $0.73{}^{+0.48}_{-0.33}\big|_{\rm fit}{}^{+0.15}_{-0.21}\big|_\beta$ & $1.75{}^{+1.26}_{-0.86}\big|_{\rm fit}{}^{+0.42}_{-0.54}\big|_\beta$\\
4.00 & $2.60{}^{+0.21}_{-0.87}\big|_\beta$ & $4.54{}^{+0.92}_{-0.74}\big|_{\rm fit}{}^{+0.95}_{-1.69}\big|_\beta$ & $7.15{}^{+2.18}_{-1.72}\big|_{\rm fit}{}^{+2.02}_{-2.81}\big|_\beta$\\
4.50 & $4.78{}^{+1.06}_{-1.93}\big|_\beta$ & $6.52{}^{+0.73}_{-0.61}\big|_{\rm fit}{}^{+2.10}_{-2.80}\big|_\beta$ & $8.96{}^{+1.70}_{-1.40}\big|_{\rm fit}{}^{+3.59}_{-4.05}\big|_\beta$\\
5.00 & $5.23{}^{+1.58}_{-2.24}\big|_\beta$ & $6.49{}^{+0.53}_{-0.45}\big|_{\rm fit}{}^{+2.53}_{-2.92}\big|_\beta$ & $8.34{}^{+1.22}_{-1.02}\big|_{\rm fit}{}^{+3.95}_{-3.94}\big|_\beta$\\
\bottomrule
\end{tabular}}
\end{table}

Inspection of Fig.~\ref{fig:componentsMB} reveals two prominent dynamical features of the reaction in the near-threshold region. First, the isolated contributions from the individual low-lying resonances $N(1535)$ and $N(1650)$ each reach several nanobarns, visibly exceeding the coherent total cross section. This behavior originates from strong destructive interference between these two $S_{11}$ states: because their relative phase is constrained by the decay data to $\phi_{1650}-\phi_{1535}\approx -155^\circ$, their coherent superposition substantially suppresses the net production rate relative to their large isolated intensities. Second, despite this destructive cancellation among the resonances, the coherent production rate remains substantially enhanced relative to the conventional pure nucleon-exchange contribution. Specifically, at $\sqrt{s}=3.60\GeV$ ($\beta=4$), the central predicted Born total cross sections reach $0.73^{+0.48}_{-0.33}\,\mathrm{nb}$ for Solution~C and $1.75^{+1.26}_{-0.86}\,\mathrm{nb}$ for Solution~D, whereas the isolated ground-state nucleon exchange yields only $0.0715\,\mathrm{nb}$.

This pronounced disparity originates from the interplay of distinct Dirac structures and reaction kinematics. Applying the Dirac equations for on-shell external nucleons, $\slsh{p}_a u(p_a)=m_N u(p_a)$ and $\bar v(p_b)\slsh{p}_b=-m_N \bar v(p_b)$, allows the nucleon-exchange amplitude in Eq.~\eqref{eq:crossN} to be expressed in the compact form:

\begin{equation}
 \amp_N^p = C_N\left[
 \frac{F_N(t)}{t-m_N^2} - \frac{F_N(u)}{u-m_N^2}
 \right]\bar v(p_b)\slsh{k}u(p_a).\label{eq:nucleon-zero}
\end{equation}

Since $t - u = 4 p_i p_f \cos\theta$, the small final-state momentum $p_f$ near threshold keeps $t$ and $u$ close ($t\approx u$). Owing to the relative minus sign between the $t$- and $u$-channel terms in Eq.~\eqref{eq:nucleon-zero}, a severe dynamical cancellation occurs between the two exchange diagrams, leading to a strong suppression of the nucleon pole near threshold. For the negative-parity $S_{11}$ resonances ($R_-$), the scalar-type vertices in Eq.~\eqref{eq:crossS} yield an additive ($+$) combination of $t$- and $u$-channel propagators that is immune to this threshold cancellation. Consequently, coherent resonance exchanges govern the near-threshold dynamics and enhance the production cross section.

\begin{figure}[!htbp]\centering
\includegraphics[width=.7\columnwidth]{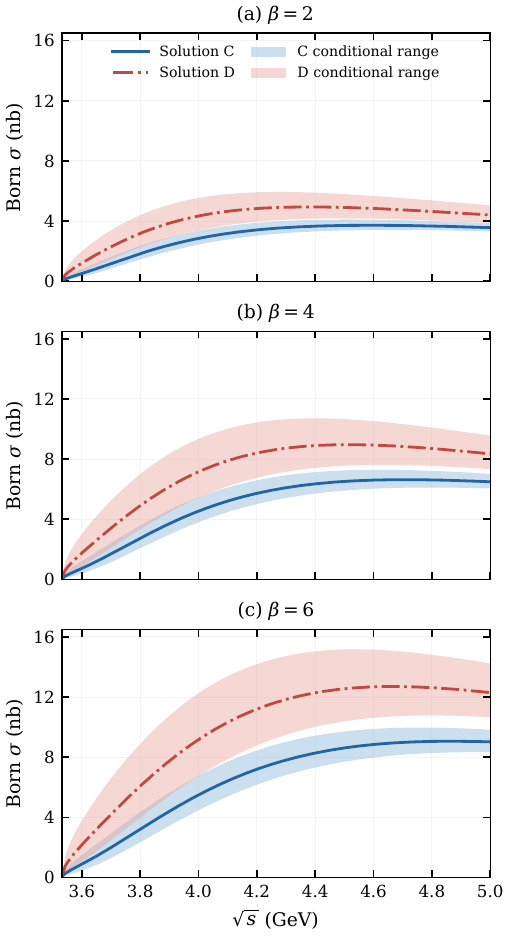}
\caption{Total Born cross sections for $p\bar{p}\to\eta_c\eta$ as a function of the center-of-mass energy $\sqrt{s}$ evaluated at three representative cutoff scales: (a) $\beta=2$, (b) $\beta=4$, and (c) $\beta=6$. Solid blue curves and dash-dotted red curves denote the central predictions for Solution~C and Solution~D, respectively. Shaded bands represent the corresponding conditional fit uncertainties at $\Delta\chi^2_{\rm comp}=1$ obtained at each fixed $\beta$.}\label{fig:productionMB}
\end{figure}

To examine the sensitivity of the production cross sections to the form-factor cutoff and compare the predictions of the two decay normalization solutions under common cutoff conditions, Fig.~\ref{fig:productionMB} displays the Born total cross sections for Solutions~C and D at three representative values: $\beta=2$, $4$, and $6$. As shown in Fig.~\ref{fig:productionMB}, the absolute scale of the cross section depends strongly on $\beta$, particularly at higher collision energies. Increasing $\beta$ systematically enhances the overall magnitude; for instance, at $\sqrt{s}=5.00\GeV$, the central prediction for Solution~C increases from $3.57$ to $9.02\,\mathrm{nb}$ as $\beta$ increases from $2$ to $6$, while Solution~D rises from $4.40$ to $12.29\,\mathrm{nb}$. The broad maxima and plateau-like features also shift in energy, showing that $\beta$ modifies the energy dependence of the off-shell suppression. Across all three cutoffs, the central cross section for Solution~D remains consistently higher than that for Solution~C.

The degree of separation between the two solutions relative to their parameter uncertainties exhibits an important energy dependence. At $\sqrt{s}=3.60\GeV$, the central values differ by a factor of $\sigma_D/\sigma_C\approx 2.4$ across all three cutoffs. The substantial fractional parameter uncertainties near threshold lead to overlapping conditional fit ranges for the two solutions (e.g., at $\beta=4$, $[0.39,\,1.21]\,\mathrm{nb}$ for Solution~C versus $[0.89,\,3.01]\,\mathrm{nb}$ for Solution~D). This overlap shows that the parameter uncertainties propagated from the present decay fits limit near-threshold discrimination between the two solutions.%

\begin{figure*}[htb]\centering
\includegraphics[width=.66\textwidth]{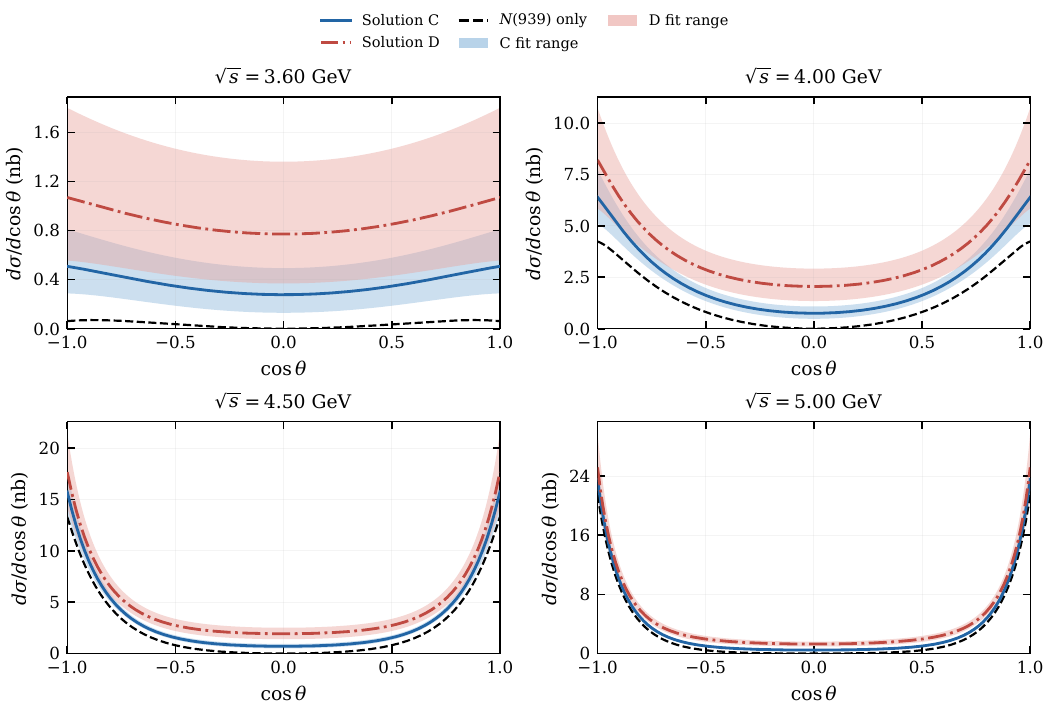}
\caption{Absolute differential cross sections $\dd\sigma/\dd\cos\theta$ (in nb) at $\beta=4$ for Solution~C (solid blue), Solution~D (dash-dotted red), and the nucleon-pole reference (dashed black) at four representative center-of-mass energies. Blue and red shaded bands denote nonlinear conditional fit uncertainties at $\Delta\chi^2_{\rm comp}=1$.}\label{fig:angleMB}
\end{figure*}

Toward higher energies, although the relative difference between the central predictions decreases, the fractional conditional fit uncertainties shrink markedly. At $\sqrt{s}=5.00\GeV$, the conditional ranges for Solutions~C and D become separated for $\beta=4$ ($[6.04,\,7.02]\,\mathrm{nb}$ vs $[7.33,\,9.56]\,\mathrm{nb}$) and $\beta=6$, while for $\beta=2$ they remain only narrowly separated ($[3.32,\,3.86]\,\mathrm{nb}$ vs $[3.88,\,5.05]\,\mathrm{nb}$). This energy-dependent behavior is governed by the evolving interplay between the fixed ground-state nucleon reference and the resonance amplitudes. In the input decay data, the branching fractions differ by a factor of $\mathcal B_D/\mathcal B_C\simeq 2.69$. The ground-state nucleon coupling $C_N$ is fixed independently from $\eta_c\to p\bar{p}$ and is common to both solutions. The fitted resonance coupling products increase by approximately a factor of $1.64$ from Solution~C to Solution~D, while their central phases remain close (see Table~\ref{tab:MB}). The coherent production amplitudes can therefore be approximated as

\begin{equation}
\begin{gathered}
 \amp_C=\amp_N+\amp_R,\quad
 \amp_D\simeq\amp_N+\lambda\amp_R,\\
 \lambda\simeq\sqrt{\mathcal B_D/\mathcal B_C}\simeq 1.64,\label{eq:scaling-amp}
\end{gathered}
\end{equation}
where $\amp_R$ is the coherent sum of the four resonance amplitudes. Squaring the coherent amplitudes yields

\begin{equation}
\begin{gathered}
 \sigma_C=\sigma_N+\sigma_R+\sigma_{NR}^{\rm int}, \quad
 \sigma_D\simeq\sigma_N+\lambda^2\sigma_R+\lambda\sigma_{NR}^{\rm int}.\label{eq:scaling}
\end{gathered}
\end{equation}
In our numerical calculations, the central ratio $\sigma_D/\sigma_C$ decreases monotonically from approximately $2.4$ at $\sqrt{s}=3.60\GeV$ to approximately $1.24$--$1.36$ at $5.00\GeV$ ($1.29$ at the reference cutoff $\beta=4$). This energy evolution can be understood qualitatively from the decomposition in Eq.~\eqref{eq:scaling}: as the collision energy increases, the rapid growth of the unscaled nucleon contribution $\sigma_N$ increases its relative weight in the total cross section, progressively diluting the contribution of the scaled resonance term $\sigma_R$ and diminishing the relative sensitivity of the total rate to the resonance normalization.

Figure~\ref{fig:productionMB} illustrates both the sensitivity
to the common cutoff parameter and the limitations
of using total cross
sections to distinguish
Solutions~C and D.

Near threshold, the sizable relative difference between the central
predictions is accompanied by overlapping conditional fit ranges,
limiting the discrimination supported by the present decay constraints.
At $\sqrt{s}=5.00\GeV$, the conditional ranges are separated at all
three representative fixed cutoffs considered here, although the
relative difference between the central predictions is smaller.
This behavior reflects the competition between the decreasing relative
difference of the central cross sections and the contraction of their
conditional fit ranges with increasing energy. At the upper end of
the energy interval studied, this competition results in a modest
separation between the conditional predictions at each common cutoff.
Given the present decay constraints and the uncertainties of a future measurement, excluding either radiative-interference solution with total cross sections alone may be difficult.

\subsection{Differential angular distributions}\label{subsec:angular-distributions}

The absolute differential cross sections $\dd\sigma/\dd\cos\theta$ for $p\bar{p}\to\eta_c\eta$ at $\sqrt{s}=3.60$, $4.00$, $4.50$, and $5.00\GeV$ are presented in Fig.~\ref{fig:angleMB}. The angular distributions provide complementary information on competing exchange mechanisms and clarify the kinematic domains where intermediate resonance contributions are most pronounced.

As derived in Eq.~\eqref{eq:nucleon-zero}, the pure nucleon-exchange amplitude exhibits an exact dynamical zero at the central scattering angle $\theta=90^\circ$ ($\cos\theta=0$). In the center-of-mass frame, $\theta=90^\circ$ strictly enforces $t=u$ across all collision energies. Therefore, the central-angle node is an exact dynamical consequence of the pseudoscalar nucleon coupling structure and the coherent cancellation between the $t$- and $u$-channel exchange diagrams. Resonance exchanges do not share this cancellation. For negative-parity states ($R_-$), the vertex operator at $t=u=z$ reduces to $2C_{R_-}(m_N+m_R)F_R(z)/(z-m_R^2)\,\bar v(p_b)u(p_a)$, which remains non-zero. At this angle, $\amp_N^p=0$ also eliminates the interference between the nucleon pole and the resonance amplitudes. Within the present framework, the differential cross section at exact $\theta=90^\circ$ is generated entirely by the coherent sum of resonance exchanges, $\dd\sigma/\dd\cos\theta|_{\theta=90^\circ}\propto |\amp_R|^2$. In actual experiments with finite angular binning around $90^\circ$, the integrated yield necessarily includes non-zero nucleon contributions away from the exact node. Nevertheless, comparison with the corresponding bin-integrated nucleon prediction provides an experimental test of the adopted pseudoscalar nucleon-exchange mechanism. While this filled node naturally arises from the intermediate resonance mechanism formulated here, it serves more broadly as evidence for dynamics beyond the pure nucleon pole rather than a unique identification of the specific resonance spectrum.

With increasing collision energy, the differential cross sections develop pronounced forward ($\cos\theta\to 1$) and backward ($\cos\theta\to -1$) peaks, characteristic of peripheral $t$- and $u$-channel exchanges. While nucleon exchange acquires increasing relative importance in the integrated rate at higher energies, the central-angle region remains sensitive to resonance contributions, as the nucleon component remains dynamically suppressed near $\cos\theta=0$. At the reference cutoff $\beta=4$, the central-angle differential rates retain a marked difference between the two solutions even when the relative difference between their integrated rates decreases. At $\sqrt{s}=5.00\GeV$, the central predictions at $\cos\theta=0$ are $0.470$ and $1.280\,\mathrm{nb}$ for Solutions~C and D, respectively, giving a ratio of approximately $2.72$ compared with $1.29$ for the integrated cross sections. The corresponding conditional ranges are $[0.346,\,0.603]\,\mathrm{nb}$ and $[0.961,\,1.617]\,\mathrm{nb}$. Thus, central angles retain sensitivity to the resonance normalization without dilution by the nucleon term at the exact node. These fixed-cutoff comparisons motivate complementary angular measurements.

To compare angular shapes independently of the overall normalization, we examine the normalized angular distribution function:
\begin{equation}
\begin{gathered}
 f_s(\cos\theta)=\frac{1}{\sigma(s)}\frac{\dd\sigma(s)}{\dd\cos\theta}, \quad
 \int_{-1}^{1}f_s(\cos\theta)\,\dd\cos\theta=1.\label{eq:moments}
\end{gathered}
\end{equation}

\begin{figure}[!thbp]\centering
\includegraphics[width=.8\columnwidth]{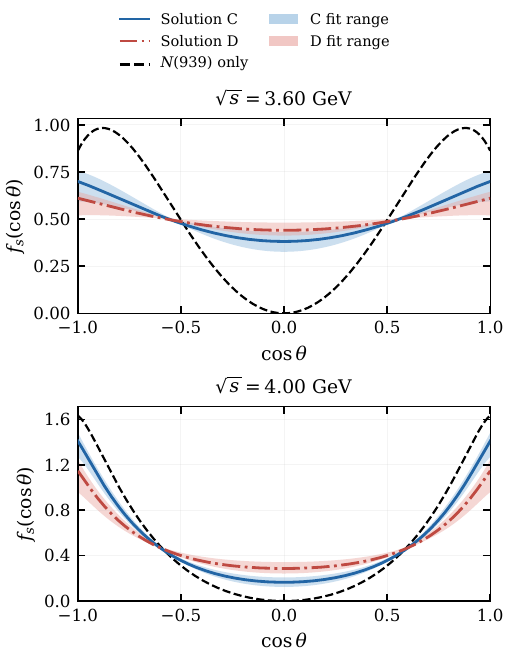}
\caption{Normalized angular distributions $f_s(\cos\theta)$ at $\sqrt{s}=3.60$ and $4.00\GeV$ for $\beta=4$.}\label{fig:normalizedMB}
\end{figure}

The normalized distributions at $\sqrt{s}=3.60$ and $4.00\GeV$ are displayed in Fig.~\ref{fig:normalizedMB}. The coherent distributions exhibit a nonvanishing central region and a reduced central suppression relative to the nucleon-only reference, which vanishes at $\cos\theta=0$. At $\sqrt{s}=3.60\GeV$, Solution~D yields a noticeably flatter angular distribution than Solution~C due to its larger relative resonance weight, and discernible curvature differences persist at $4.00\GeV$. At $4.00\GeV$ and $\cos\theta=0$, the conditional ranges are $[0.125,\,0.208]$ for Solution~C and $[0.243,\,0.343]$ for Solution~D, whereas the central-angle ranges overlap at $3.60\GeV$. 

We emphasize that the cross sections presented here are evaluated at the tree level (Born approximation). In physical antiproton--proton annihilation, strong initial-state interactions (ISI) can attenuate exclusive cross sections through absorptive multi-channel transitions~\cite{Haidenbauer:2009ad,Haidenbauer:2014rva}. The magnitude of ISI suppression depends on the partial-wave structure and short-range dynamics of the final state, so absorption factors from other charmonium channels cannot be transferred trivially to $\eta_c\eta$. These calculations provide hadronic input for future distorted-wave Born approximation (DWBA) or full coupled-channel analyses.

\section{Summary}\label{sec:summary}

We have carried out a study of the associated charmonium production reaction $p\bar{p}\to\eta_c\eta$ from threshold up to $\sqrt{s}=5.00\GeV$ within an effective Lagrangian framework. Our analysis is constrained by the first observation of the three-body decay $\eta_c\to p\bar{p}\eta$ by the BESIII Collaboration~\cite{BESIII:2026kdm}. In our model, the reaction is assumed to proceed via the coherent $t$- and $u$-channel exchanges of the ground-state nucleon $N(939)$ and four established spin-$1/2$ resonances: $N(1535)$, $N(1650)$, $N(1895)$, and $N(1710)$. The selection of these intermediate states is guided by physical criteria: prioritizing low-lying states with smaller off-shell virtualities, low partial waves to minimize centrifugal barrier suppression, and sizeable empirical branching fractions to the $N\eta$ decay channel.

The ground-state nucleon coupling was fixed independently from the empirical $\eta NN$ coupling and the $\eta_c\to p\bar{p}$ decay width. The effective coupling products and relative phases of the four nucleon resonances were determined through simultaneous fits to the BESIII invariant-mass projections ($M_{p\eta}$, $M_{\bar{p}\eta}$, and $M_{p\bar{p}}$) for both constructive (Solution~C) and destructive (Solution~D) radiative-interference normalizations. The decay fits describe the low-mass enhancement near $1.53\GeV$ through coherent $N(1535)$--$N(1650)$ interplay and destructive interference among the retained amplitudes, with their relative phase tightly constrained to $\phi_{1650}-\phi_{1535}\approx -155^\circ$.

Extending these decay-constrained amplitudes to the crossed production channel via the constant-phase continuation prescription reveals several key dynamical features within the adopted Born framework. First, coherent resonance exchanges govern the near-threshold production dynamics and generate a prominent cross-section enhancement. At $\sqrt{s}=3.60\GeV$ ($\beta=4$), the central predicted total Born cross sections reach $0.73^{+0.48}_{-0.33}\,\mathrm{nb}$ for Solution~C and $1.75^{+1.26}_{-0.86}\,\mathrm{nb}$ for Solution~D, exceeding the isolated nucleon-exchange contribution ($0.0715\,\mathrm{nb}$) by more than an order of magnitude (with conditional lower bounds remaining several times higher), while a substantial enhancement persists across the entire cutoff scan $2\leq\beta\leq6$. Physically, this behavior originates from the scalar-type Dirac structures of the negative-parity $S_{11}$ states, which yield an additive combination of $t$- and $u$-channel amplitudes that is immune to the dynamical cancellation suppressing the pseudoscalar nucleon pole near threshold.

Second, the energy evolution of the total cross sections reflects the changing dynamical interplay between nucleon and resonance exchanges, illustrating both the potential and the limitations of testing the two radiative-interference solutions. In our framework, the rapid growth of the unscaled nucleon contribution progressively increases its relative weight in the total rate, thereby diluting the scaling resonance terms. Consequently, the central cross-section ratio $\sigma_D/\sigma_C$ decreases monotonically from $\sim 2.4$ at $\sqrt{s}=3.60\GeV$ to $1.24$--$1.36$ at $5.00\GeV$ ($1.29$ at the reference cutoff $\beta=4$). Near threshold, the relative difference between the central predictions is sizable, but the conditional fit ranges overlap, limiting discrimination between the two solutions under the present decay constraints. At $\sqrt{s}=5.00\GeV$, the conditional ranges are separated at the three representative fixed cutoffs considered here; however, the relative difference between the central predictions is smaller and the absolute rates depend strongly on the cutoff parameter $\beta$. Once experimental measurement uncertainties are taken into account, cleanly excluding either solution using production total cross sections alone may be challenging with the present decay constraints.

Third, the differential angular distributions provide characteristic angular signatures of intermediate resonance dynamics. Owing to the pseudoscalar nucleon coupling structure, the pure nucleon-exchange amplitude undergoes an exact dynamical cancellation at $\theta=90^\circ$ ($t=u$), producing an exact node across all collision energies. In our model, coherent resonance exchanges fill this node, with both the isolated nucleon term and the nucleon--resonance interference vanishing identically at $90^\circ$. The central-angle distribution therefore provides a sensitive test of the pseudoscalar nucleon-exchange mechanism adopted here. At the exact central angle, the absence of nucleon dilution preserves sensitivity to the resonance normalization (for instance, at $\sqrt{s}=5.00\GeV$ and $\beta=4$, the ratio $(\dd\sigma/\dd\cos\theta)_D / (\dd\sigma/\dd\cos\theta)_C$ at $90^\circ$ remains $\sim 2.72$, well above the integrated cross-section ratio of $1.29$), motivating complementary angular measurements alongside total cross sections.

These results demonstrate that intermediate nucleon resonances substantially reshape both the absolute scale and the angular characteristics of the $p\bar{p}\to\eta_c\eta$ reaction within the adopted framework. Future antiproton experiments, such as those planned by the $\bar{\text{P}}\text{ANDA}$ Collaboration at FAIR, could test these predicted excitation functions and angular profiles, thereby providing valuable insights into charmonium--nucleon dynamics and baryon spectroscopy.

\begin{acknowledgments}

This work is supported by the Natural Science Foundation of Gansu Province (No. 26RCKA012 and No. 25JRRA799), the National Natural Science Foundation of China under Grants No. 12335001 and No. 12247101, the ``111 Center" under Grant No. B20063, the Fundamental Research Funds for the Central Universities (lzujbky-2023-stlt01), and Lanzhou City High-Level Talent Funding. The authors used ChatGPT to assist with literature searches, numerical calculations, and figure preparation, and Gemini for language editing. The authors reviewed the output and take responsibility for the content.
\end{acknowledgments}

\section*{Data Availability}
The experimental inputs were extracted from the invariant-mass distributions and normalized sideband backgrounds published by the BESIII Collaboration~\cite{BESIII:2026kdm}. The data supporting this study are available from the corresponding author upon reasonable request.

\appendix
\section{Experimental constraints, fitting procedure, and uncertainty analysis}\label{app:experimental-input}\label{app:uncertainties}

To constrain the model parameters, we utilize the experimental invariant-mass distributions of candidate events in the $\eta_c$ signal region together with the normalized sideband backgrounds from Fig.~4 of Ref.~\cite{BESIII:2026kdm}. The published experimental spectra for the $p\eta$, $\bar p\eta$, and $p\bar p$ subsystems comprise 72, 73, and 72 bins, respectively, with a uniform bin width of $\Delta m_b=8\MeV$. In the BESIII measurement, quantum interference between the resonant $M1$ transition $\psi(3686)\to\gamma\eta_c$ and the non-resonant continuum $\psi(3686)\to\gamma p\bar{p}\eta$ yields two distinct branching-fraction solutions:
\begin{equation}
\begin{gathered}
 \text{Solution C (constructive):}\\\mathcal B_C=(0.90\pm0.04\pm0.21\pm0.13)\times10^{-3},\\
 \text{Solution D (destructive):}\\\mathcal B_D=(2.42\pm0.07\pm0.48\pm0.34)\times10^{-3}.
\end{gathered}\label{eq:branches}
\end{equation}
These solutions provide alternative absolute normalizations for the $\eta_c\to p\bar{p}\eta$ decay rate. We note that ``C'' and ``D'' refer exclusively to interference in the primary radiative transition; relative phases among intermediate hadronic amplitudes are determined independently by our fits.

For each two-body projection $\alpha\in\{p\eta, \bar p\eta, p\bar p\}$ and mass bin $b$, let $D_{\alpha b}$ and $B_{\alpha b}$ denote the observed signal-window counts and scaled background events, respectively. The background-subtracted event yield is
\begin{equation}
\begin{gathered}
 Y_{\alpha b}=D_{\alpha b}-B_{\alpha b}, \quad
 S_\alpha=\sum_b Y_{\alpha b},\\
 f_{\alpha b}=\frac{Y_{\alpha b}}{S_\alpha\Delta m_b},\quad
 \sum_b f_{\alpha b}\Delta m_b=1,\label{eq:shape}
\end{gathered}
\end{equation}
where $S_\alpha$ is the total net signal yield and $f_{\alpha b}$ is the normalized event density. We emphasize that this treatment adopts an effective common-shape approximation derived directly from the background-subtracted candidate distributions for both Solutions~C and D. While detector acceptance is assumed to vary smoothly across phase space, explicit detector resolution unfolding is not performed. Furthermore, empirical sideband subtraction does not account for possible coherent quantum interference between the non-resonant continuum and the $\eta_c$ signal within the signal window. Consequently, the fitted parameters should be interpreted as constrained by these effective candidate spectra rather than an unfolded, pure decay Dalitz distribution.

To confront the theoretical decay amplitudes with data, experimental counts are converted into absolute differential widths and bin-integrated partial widths for each solution $r\in\{C, D\}$:
\begin{align}
 \left(\frac{\dd\Gamma_\alpha}{\dd m}\right)^{(r)}_b
   &=\Gamma_{\eta_c}\mathcal B_r f_{\alpha b},\nonumber\\
 y_{\alpha b}^{(r)}&=a_\alpha^{(r)}Y_{\alpha b},\qquad
 a_\alpha^{(r)}=\frac{\Gamma_{\eta_c}\mathcal B_r}{S_\alpha},\nonumber\\
 \epsilon_{\alpha b}^{(r)}&=a_\alpha^{(r)}
 \max\left[\frac{\epsilon^+_{D,\alpha b}+\epsilon^-_{D,\alpha b}}2,1\right],\label{eq:data}
\end{align}
where $\Gamma_{\eta_c}=30.5\MeV$ is the total width of $\eta_c$, and $\epsilon^\pm_{D,\alpha b}$ are experimental Poisson count errors, with an error floor of one event imposed for statistical stability. Under the common-shape approximation, the absolute differential inputs obey
\begin{equation}
 \frac{(\dd\Gamma_\alpha/\dd m)^{(D)}_b}
 {(\dd\Gamma_\alpha/\dd m)^{(C)}_b}
 =\frac{\mathcal B_D}{\mathcal B_C}\simeq2.69\qquad(f_{\alpha b}\ne0).\label{eq:datanorm}
\end{equation}

The composite objective function minimized during fitting is
\begin{equation}
 \chi^2_{\rm comp}=\sum_{\alpha=p\eta,\bar p\eta,p\bar p}\sum_b
 \left[\frac{\Delta\Gamma_{\alpha b}^{\rm th}-y_{\alpha b}^{(r)}}
 {\epsilon_{\alpha b}^{(r)}}\right]^2.\label{eq:chi}
\end{equation}
Theoretical partial widths $\Delta\Gamma_{\alpha b}^{\rm th}$ are integrated over the physical phase-space boundary. Summing over the three projections gives 217 residual terms for eight real parameters (four coupling products and four phases), or 209 nominal degrees of freedom. In the absence of an experimental covariance matrix between the different projections, Eq.~\eqref{eq:chi} adopts standard diagonal weighting. The external inputs—background subtraction, normalization sums, branching fractions, hadron masses, and widths—are held fixed during parameter optimization, and the resulting parameter uncertainties represent conditional model sensitivities derived from profiling the composite $\chi^2$ objective function.

Parameter uncertainties compiled in Table~\ref{tab:MB} are evaluated at the reference cutoff $\beta=4$ using HESSE for the local Hessian and MINOS for asymmetric one-parameter profile ranges at $\Delta\chi^2_{\rm comp}=1$~\cite{James:1975dr}. The conditional uncertainty bands shown in Figs.~\ref{fig:componentsMB}, \ref{fig:angleMB}, and \ref{fig:normalizedMB} are evaluated at this reference cutoff $\beta=4$. In contrast, for the comparison in Fig.~\ref{fig:productionMB}, the conditional profile bands are computed independently for each panel at its corresponding fixed cutoff $\beta \in \{2, 4, 6\}$: at each $\beta$, the parameters are refitted to find the local optimum $\widehat{\boldsymbol\theta}(\beta)$, and the observable $O$ is extremized over the corresponding conditional profile region
\begingroup
\begin{equation}
\begin{aligned}
 \Omega_1(\beta)=\{\boldsymbol\theta:\;&\chi^2_{\rm comp}(\boldsymbol\theta;\beta)-\chi^2_{\rm comp}(\widehat{\boldsymbol\theta}(\beta);\beta)\leq1\},
 \\O_-=\min_{\Omega_1(\beta)}O,&O_+=\max_{\Omega_1(\beta)}O.\label{eq:profile-observable}
\end{aligned}
\end{equation}
\endgroup
These bands reflect conditional parameter sensitivities under fixed external inputs (branching ratios, masses, and widths) and effective experimental inputs, rather than fully calibrated absolute confidence intervals.
For each profiled observable, the complete coherent amplitude and all parameter correlations are retained. For normalized angular distributions, the numerator and denominator are evaluated simultaneously at each parameter point. Cutoff scans are performed over $\mathcal G_\beta=\{2.0,2.1,\ldots,6.0\}\cup\mathcal G_{\rm loc}$. At each $\beta_b$, the decay parameters are refitted to obtain the optimized parameter set $\widehat{\boldsymbol\theta}(\beta_b)$. Relative to the reference cutoff $\beta=4$, the cutoff sensitivity envelopes are defined by
\begin{align}
\delta^-_\beta O&=O(\widehat{\boldsymbol\theta}(4),4)\nonumber-\min_{\beta_b\in\mathcal G_\beta}O(\widehat{\boldsymbol\theta}(\beta_b),\beta_b),\nonumber\\
\delta^+_\beta O&=\max_{\beta_b\in\mathcal G_\beta}O(\widehat{\boldsymbol\theta}(\beta_b),\beta_b)\nonumber-O(\widehat{\boldsymbol\theta}(4),4).\label{eq:beta-range}
\end{align}
These sensitivity envelopes quantify form-factor systematics independently of the conditional parameter fit uncertainties, and are therefore quoted separately rather than added in quadrature.

\bibliographystyle{apsrev4-2}
\bibliography{references}

\end{document}